\documentclass[preprint2]{aastex}
\usepackage{graphicx}
\usepackage{natbib}

\shorttitle{X-ray and radio observations of YSOs in NGC~1333 and IC~348}
\shortauthors{Forbrich, Osten, \& Wolk}

\begin{document}

\title{Simultaneous X-ray and radio observations of Young Stellar Objects in NGC~1333 and IC~348}

\author{Jan Forbrich\altaffilmark{1}, Rachel Osten\altaffilmark{2}, \& Scott J. Wolk\altaffilmark{1}}

\altaffiltext{1}{Harvard-Smithsonian Center for Astrophysics, 60 Garden Street, Cambridge, MA 02138, USA}
\altaffiltext{2}{Space Telescope Science Institute, 3700 San Martin Drive, Baltimore, MD 21218, USA}

\begin{abstract}
Young Stellar Objects (YSOs) and in particular protostars are known to show a variety of high-energy processes. Observations in the X-ray and centimetric radio wavelength ranges are thought to constrain some of these processes, e.g., coronal-type magnetic activity. There is a well-known empirical correlation of radio and X-ray luminosities in active stars, the so-called G\"udel-Benz relation. Previous evidence whether YSOs are compatible with this relation remains inconclusive for the earliest evolutionary stages. The main difficulty is that due to the extreme variability of these sources, simultaneous observations are essential. Until now, only few YSOs and only a handful of protostars have been observed simultaneously in the X-ray and radio range. To expand the sample, we have obtained such observations of two young clusters rich in protostars, NGC 1333 and IC 348. While the absolute sensitivity is lower for these regions than for more nearby clusters like CrA, we find that even in deep continuum observations carried out with the NRAO Very Large Array, the radio detection fraction for protostars in these clusters is much lower than the X-ray detection fraction. Very few YSOs are detected in both bands, and we find the radio and X-ray populations among YSOs to be largely distinct. We combine these new results with previous simultaneous \textit{Chandra} and VLA observations of star-forming regions and find that YSOs with detections in both bands appear to be offset toward higher radio luminosities for given X-ray luminosities when compared to the G\"udel-Benz relation, although even in this sensitive dataset most sources are too weak for the radio detections to provide information on the emission processes. The considerably improved sensitivity of the Expanded Very Large Array will provide a better census of the YSO radio population as well as better constraints on the emission mechanisms.
\end{abstract}


\section{Introduction}

Young Stellar Objects (YSOs) have been known for some time to be observable in X-ray and radio emission, tracers of high-energy processes \citep[e.g., ][]{fei99}. Briefly, thermal X-ray emission is thought to be produced by hot plasma heated in coronal-type activity while non-thermal radio emission traces magnetic fields that may be cospatial to and confining the X-ray emitting plasma. YSOs are typically grouped into classes, starting with class 0/I protostars as the earliest stages, and progressing through disk-dominated class II objects to basically diskless class III sources, also known as weak-line T Tauri stars. These objects are progressively less embedded, and the importance of first the protostellar envelope and then the protostellar disks diminishes \citep{lad87,and93}.

There are several effects which complicate a direct comparison of the high-energy processes of YSOs and solar-like stars. Circumstellar disks, a defining characteristic of protostars, may affect the overall shape of the magnetic field, for example by providing support for large magnetic structures connecting the central object and the disk. Radio emission may be dominated by either nonthermal emission originating in magnetic fields, or thermal emission from the disk or stellar wind. The latter is usually interpreted as thermal bremsstrahlung due to ionized material while the former most frequently is (gyro)synchrotron radiation (e.g., \citealp{dul85,and96,gue02}). Note that thermal emission can easily conceal nonthermal emission from underneath due to its optical depth. \citet{and87} estimate that a spherical ionized stellar wind of only a few 10$^{-11}$\, $M_\odot$\,yr$^{-1}$ is already opaque. While it may thus seem unlikely that there is a chance to detect any nonthermal emission from protostars at all, such cases have been reported \citep{fei98,shi07}.

There are a few different methods to identify nonthermal radio sources in contrast to thermal sources. The two primary indications are polarization and negative spectral indices, as derived from the two simultaneously measured radio bands. Here, the spectral index $\alpha$ is defined as $S_\nu \propto \nu^\alpha$, and a negative spectral index $\alpha<-0.1$ is an indication of optically thin nonthermal emission (e.g., \citealp{and96}). While a two-frequency spectral index is not a very reliable indicator, the detection of polarized emission is the most unambiguous sign of non-thermal emission as observed with the VLA. However, due to the small degree of polarization (e.g., \citealp{whi92,fei98}), high S/N ratios are required to meaningfully detect or constrain such polarization. A third indication is the existence of short-term variability. 

While it is not obvious that thermal hot-plasma X-ray emission and nonthermal radio emission should have an observable underlying connection, certain types of solar flares and a variety of active stars have been shown to have correlated X-ray and (nonthermal) radio luminosities according to the so-called G\"udel-Benz (GB) relation, $L_{\rm X}/L_{\rm R} \approx 10^{15\pm1}$~Hz, extending over 10 orders of magnitude \citep{gue93,ben94,gue02}. 
The explanation which these authors suggested for the nearly linear correlation between X-ray and radio luminosities relied on a common energy reservoir for both plasma heating and particle acceleration, and a similar partitioning of energy into these two processes over 10 orders of magnitude in stellar luminosity.
For a recent summary of the X-ray--radio correlation in cool stars, see \citet{for11}. 

The early studies on correlated X-ray and radio luminosities of different object classes included some weak-line T Tauri stars (class III sources) as powerful radio and X-ray sources, even though the underlying observations were non-simultaneous. It is therefore interesting to find out whether YSOs in earlier evolutionary stages, particularly class~I protostars -- strong X-ray and radio sources in their own right -- are compatible with the GB relation. Since YSOs are highly variable, this question is preferably addressed by simultaneous X-ray and radio observations. Until now, there have been few such observations of YSOs. V773 Tau, a T Tauri star, became the first YSO observed in this way, and radio variability exceeding the X-ray variability was found \citep{fei94,gue00}. \citet{bow03} reported the serendipitous simultaneous radio and X-ray observation of a spectacular radio flare toward a weak-line T Tauri star in Orion (GMR-A). In the first study targeting a subset of an entire star-forming region, \citet{gag04} observed the $\rho$ Oph cloud complex with \textit{Chandra} and the NRAO Very Large Array. Several T Tauri stars were detected in both wavelength ranges, but no clear X-ray--radio correlation was found. Subsequently, \citet{for07} succeeded in detecting several class I protostars in the CrA star-forming region simultaneously in the X-ray and radio wavelength ranges. Among the YSOs in CrA detected in both bands, most are compatible with the GB relation. Most recently, \citet{ost09} report the detections in both bands of several T Tauri stars in the LkH$\alpha$\,101 cluster. Here, no clear X-ray--radio correlation was found.

In an attempt to observe more protostars simultaneously in the X-ray and radio wavelength ranges, we have observed selected regions in two more young clusters, IC~348 and NGC~1333. Previously published observations carried out with the \textit{Spitzer Space Telescope} serve as an excellent road map to locate YSOs in early evolutionary stages in these clusters. Two-frequency radio observations help us in potentially discriminating between nonthermal and thermal radio sources, based on the value of the radio spectral indices. We will discuss the results in the context of previous simultaneous radio and X-ray observations of YSOs.

\subsection{The target regions: IC 348 and NGC 1333}

Located in the Perseus Molecular Cloud, both IC~348 and NGC~1333 are prominent nearby star-forming regions (for recent reviews, see \citealp{her08} and \citealp{wal08}, respectively, and \citealp{bal08}). The most accurate distance has been determined toward NGC~1333 by measuring the parallax of the bright maser source SVS 13 \citep{hir08}, yielding 235$\pm$18~pc. IC~348 is slightly more distant at about 260~pc \citep{lom10}. The star formation activity in both regions has been surveyed extensively. Most relevant for our purposes are studies using the \textit{Spitzer Space Telescope} to identify and classify YSOs in early evolutionary stages by their infrared excess emission. \citet{lad06} analyzed $\sim$300 previously known members of IC 348 and classified them by their disk properties, distinguishing stellar (photospheric), anemic, and thick-disk spectral energy distributions (SEDs). The main basis for this classification is the dereddened \textit{Spitzer}-IRAC slope. In a follow-up paper, \citet{mue07} identified and analyzed new cluster members, including embedded class~I protostars. Their class~II category corresponds to the thick disks from \citeauthor{lad06}, even though the spectral indices of some thick disk sources from \citet{lad06} qualify as class I sources according to \citet{mue07}. To use a consistent classification, we follow the spectral index limits in \citet{mue07} without separating out flat-spectrum sources. The class~III objects in \citet{mue07} correspond to the anemic and photospheric sources in \citet{lad06}. The NGC~1333 region was similarly surveyed by \citet{gut08} who employ a three-phase technique to identify YSOs. Our focus here is on the youngest evolutionary stages. This is due both to scientific interest and the fact that in both regions previous surveys have been most complete for class~I and class~II sources. However, our study also reveals class~III sources. Since the \textit{Spitzer} studies cannot distinguish class 0 and class I sources, we additionally use the sources identified by \citet{eno09} in Perseus to find the class 0 sources among the \textit{Spitzer} class I sources.
Compared to star-forming regions that have previously been targets of simultaneous X-ray and radio observations (see above), IC~348 and NGC~1333 are about twice as distant as Taurus, $\rho$~Oph, and CrA, but at less than half the distance of LkH$\alpha$~101.

\subsection{Previous radio observations}

Both IC~348 and NGC~1333 have been observed with the NRAO\footnote{The National Radio Astronomy Observatory is a facility of the National Science Foundation operated under cooperative agreement by Associated Universities, Inc.} Very Large Array (VLA) prior to our own observations. Our region of interest in IC~348 has been observed by \citet{avi01} who present a deep radio dataset centered on HH\,211. Their concatenated dataset has a quoted rms noise sensitivity of 12~$\mu$Jy and consists of a short C array observation obtained in 1994 and 9.7~h of on-source time using 11 VLA antennas in 1999. The phase center of these observations lies $\sim9'$ off our position 1 (see below) and $\sim2'$ off our position 2.

NGC~1333 was studied in VLA observations of SVS~13 and its surroundings. After an initial study \citep{rod97} with two observations reaching sensitivities of 30~$\mu$Jy and 20~$\mu$Jy, deep concatenated archival datasets in the 3.6~cm (X)  and 6~cm bands (C) were presented by \citep{rod99}. While these datasets contain data obtained in various configurations, a common subset of the $uv$ plane has been used for the combined data, resulting in angular resolutions of 4$''$--5$''$ in both bands. The phase centers of these observations coincide with our southern pointing (see below) whereas our northern pointing is off by $\sim7'$.

\section{Observations and data analysis}

\begin{figure*}
\begin{minipage}{0.49\linewidth}
%\vspace*{-8mm}
%\includegraphics[width=0.98\linewidth]{ic348_cxovla.eps}
\includegraphics[width=0.98\linewidth]{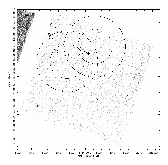}
%\caption{\label{ic348_cxovla}}
%\end{figure*}
\end{minipage}
%\begin{figure*}
\begin{minipage}{0.49\linewidth}
%\vspace*{7mm}
%\includegraphics[width=\linewidth]{ngc1333_cxovla.eps}
\includegraphics[width=\linewidth]{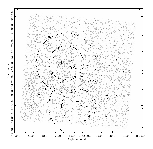}
\end{minipage}
\caption{Merged \textit{Chandra} pointings toward IC\,348 (left panel) and NGC 1333 (right panel). The eastern and western pointings in IC~348 are positions 1 and 2, respectively, and the northern and southern pointings in NGC~1333 are positions N and S. The large circles indicate the primary beam sizes for the VLA pointings, the C band having a larger primary beam than the X band. YSOs detected in both bands (from Table~\ref{tbl-radio-xray}) are marked by small circles. \label{ic348ngc1333_cxovla}}
\end{figure*}

Both IC 348 and NGC 1333 have been observed with very similar observational setups. Since the \textit{Chandra}-ACIS field of view ($17'\times17'$) is considerably larger than the VLA primary beam, we have selected two VLA pointings within the \textit{Chandra} fields (see Figure~\ref{ic348ngc1333_cxovla}). Using the VLA with two distinct subarrays allows us to record two different radio frequencies simultaneously. 
%DUP: The advantage of using two subarrays is that we obtain spectral index information to identify nonthermal radiation. 

\subsection{IC 348}
A region centered on RA 03h43m59.90s, Dec +31$^\circ$58$'$21.70$''$ was observed with \textsl{Chandra} ACIS-I in two 40~ksec observations on March 13 and on March 18, 2008 (IDs 8933 and 8944). Simultaneous radio observations were carried out with the VLA in its C configuration (project ID S9056). Two subarrays were used to cover the X and C bands simultaneously, using 13 antennas in X band and 14 antennas in C band. Two pointing positions within the ACIS-I field of view were observed alternatingly, RA 03h44m13.58s, Dec +32$^\circ$00$'$24.21$''$ (position 1) and RA 03h43m52.96s, Dec +32$^\circ$02$'$44.21$''$ (position 2). Total on-source time for each pointing and frequency is  approximately 10.0 hours. The full-width half-maximum primary beam sizes in the X and C bands are $5.3'$ and $9.6'$, respectively. An absolute flux density scale was established by an observation of 3C147 on the second day of observations. The phase calibrator 0336+323 was observed after each 20 min scan on one of the two pointing positions. The primary calibrator (3C147) was at 4.68~Jy (X-band) and 7.86~Jy (C-band), and the phase calibrator 0336+323 had bootstrapped flux densities of 1.24~Jy (X-band) and 1.68~Jy (C-band). 
%Combining both pointings, the nominal FWHM primary beam areas cover 13 class I sources in both X and C bands \citep{mue07}.

\subsection{NGC 1333}
A region centered on RA 03h29m02.00s, Dec +31$^\circ$20$'$54.00$''$ was observed with \textsl{Chandra} ACIS-I in two 40~ksec observations on July 5 and on July 11, 2006 (IDs 6436 and 6437). Simultaneous radio observations were carried out with the NRAO Very Large Array (VLA) in its B configuration (project ID S7874). Two subarrays with 13 antennas each were used to cover the X and C bands simultaneously. Two pointing positions within the ACIS-I field of view were observed alternatingly, RA 03h29m14.072s, Dec +31$^\circ$22$'$40.06$''$ (position N) and RA 03h29m04.941s, Dec +31$^\circ$15$'$54.37$''$ (position S). The total on-source time for each pointing and frequency is approximately 9.4 hours and the typical on-source scan-duration was 15~minutes. An absolute flux density scale was established by observations of 3C286 on both days of observations. The phase calibrator 0336+323, the same that was also used for IC~348, was observed after each scan on one of the two pointing positions. The primary calibrator (3C286) was at 5.20~Jy (X-band) and 7.49 (C-band), and the phase calibrator 0336+323 had bootstrapped flux densities of 1.91~Jy (X-band) and 2.28 Jy (C-band). 
%Combining both pointings, the nominal FWHM primary beam areas cover 12/28 class I sources (X-band/C-band; \citealp{gut08}). 

\subsection{Sensitivity}
Given the different sizes of the fields of view of \textit{Chandra}-ACIS and the VLA, there is a tradeoff between increasing spatial coverage by using several VLA pointings and increasing sensitivity by observing only one position. Additionally, observations at only one frequency would be more sensitive than the use of two frequencies, either by using two subarrays or by observing separately. As a compromise, we opted for a setup using two pointings and two frequency bands: a loss of sensitivity. Compared to a setup where we would maximize sensitivity by observing a single position with all antennas at one frequency, we lose a factor of 2$\cdot\sqrt{2}$=2.8 in flux density sensitivity. Primarily due to the different distances to star-forming regions, this study is not as sensitive as the $\rho$ Oph study by \citet{gag04} or the CrA study by \citet{for07}. Our observation strategy is a compromise between the one employed by \citet{gag04} who used a mosaic of VLA pointings to cover the \textit{Chandra} field and \citet{for06} who maximized sensitivity by using only one pointing position and frequency. We show a more detailed comparison of the various simultaneous \textit{Chandra} and VLA observations of star-forming regions in Table~\ref{tbl-cxo-vla}.

\begin{deluxetable}{lrrrrl}
\tablecaption{Sensitivities of simultaneous \textit{Chandra} and VLA observations of star-forming regions\label{tbl-cxo-vla}}
\tablecolumns{6}
\tablehead{
\colhead{Target} & \colhead{Date} & \colhead{$d$} & \colhead{$L_R$(5$\sigma$)}	   & \colhead{$L_X$(5\,cnts)} & \colhead{References}\\
\colhead{}       & \colhead{}	  & \colhead{(pc)}       & \colhead{(erg\,s$^{-1}$\,Hz$^{-1}$)} & \colhead{(erg\,s$^{-1}$)} & \colhead{}
}
\startdata
$\rho$ Oph&  May 2000 & 130\tablenotemark{a} & $3.2\times10^{15}$ & $7.8\times10^{26}$ & \citet{gag04}\\ % gag04: 165pc scaled to 130 pc
LkHa101   &  Mar 2005 & 510 & $1.5\times10^{16}$ & $1.5\times10^{28}$ & \citet{ost09}\\
CrA       &  Aug 2005 & 130 & $1.7\times10^{15}$ & $1.0\times10^{27}$ & \citet{for07}\tablenotemark{b}\\
NGC 1333  &  Jul 2006 & 235 & $6.9\times10^{15}$ & $3.2\times10^{27}$ & this work\\
IC 348    &  Mar 2008 & 250 & $5.7\times10^{15}$ & $3.6\times10^{27}$ & this work\\
\enddata
\tablecomments{The radio and X-ray sensitivities have been determined using the 5$\sigma$ radio detection limit in either C or X band and assuming a 5 count detection limit of an unabsorbed source with an APEC hot-plasma spectrum (with a temperature of 1~keV) in the \textit{Chandra} data, as determined using PIMMS \citep{muk93}.}
\tablenotetext{a}{\citet{gag04} assume $d=165$~pc, but see discussion in \citet{wil08}.}
\tablenotetext{b}{From simultaneous observations. For deeper radio and X-ray data, see \citet{for06} and \citet{fop07}.}
\end{deluxetable}

\subsection{Data reduction and analysis}

All VLA observations were analyzed using the NRAO Astronomical Image Processing System (AIPS), following standard procedure. After data flagging and calibration, the primary beam areas of both experiments were imaged using a robustness power of 0 for {\sl uv} weighting as a compromise between sensitivity and synthesized beam size. For the best sensitivity in subsequent source detection both observing epochs of the two clusters were combined for imaging. Source detection was performed using the AIPS task `SAD' above a S/N ratio of 5. All maps have rms noise levels near the theoretical expectation of 18~$\mu$Jy, based on the VLA Exposure Calculator\footnote{\url{http://www.vla.nrao.edu/astro/guides/exposure/calc.html}}, assuming 10 hours of on-source time at full bandwidth, effectively 86 MHz, and 13 antennas. The fitted source flux densities were corrected for primary beam attenuation and the peak flux densities from the fits are reported in the tables. Radio upper limits were determined on maps that were previously corrected for primary-beam attenuation. The synthesized beam sizes are on the order of 2.5$''$ (IC 348) and 0.8$''$ (NGC 1333) in X band and on the order of 4.0$''$ (IC 348) and 1.2$''$ (NGC 1333) in C band.

%We have employed three different methods to identify nonthermal radio sources in contrast to thermal sources. The two primary indications are circular polarization (apparent in a Stokes-\textit{V} map) and negative spectral indices, as derived from the two simultaneously measured radio bands. Here, the spectral index $\alpha$ is defined as $S_\nu \propto \nu^\alpha$, and a negative spectral index $\alpha<-0.1$ is an indication of optically thin nonthermal emission (e.g., \citealp{and96}). A third indication is the existence of variability. To this end, we have repeated imaging and source detection in each of the two epochs in both experiments.

The \textit{Chandra} X-ray observations have been reduced using the ANCHORS pipeline (AN archive of Chandra Observations of Regions of Star formation\footnote{http://cxc.harvard.edu/ANCHORS}). The X-ray observations of NGC~1333 have already been published elsewhere \citep{win10}. Full details of the ANCHORS processing are also given there. For previous X-ray results on NGC~1333, see also \citet{get02}.

%%%%%%%%%%%%%%%%%%%%%%%%%%%%%%%%%%%%%%%%%%

\section{Results}

\subsection{Radio data}

\subsubsection{IC 348}

Only five radio sources were detected in the two integrated IC 348 pointings above a significance of 5$\sigma$ (map rms 14~$\mu$Jy). In addition to using the AIPS task SAD, we have manually searched about 160 YSO positions reported by \citet{lad06} and \citet{mue07} for lower-significance detections in either the X or C-bands, if they were reasonably close to the pointing centers. This search resulted in two additional detections, VLA 2 and LRL 51. While VLA 2 has been reported as a radio source before, LRL 51 has not. Since LRL 51 is detected in both radio bands, it seems unlikely that this would be due to a chance alignment with a noise peak; see the discussion of chance alignments for NGC 1333 below.

In Table~\ref{tbl-vla-i348}, we list the total of seven radio detections in the nominal primary beam areas (FWHM) from both detection methods. Five sources can be identified with sources from the samples of \citet{lad06} and \citet{mue07}. These are HH\,211-mm, a class 0 protostar, LRL 52590 and LRL 51, class I sources, LRL 13 (class~II), as well as LRL 49, a source listed as having a stellar SED (class~III). Additionally, we note weak outflow-like emission on opposing sides of LRL 54459/54460 and LRL 57025, but these tentative detections are not listed in the source table. Source 7 does not have infrared or optical counterparts; it is thus probably extragalactic.

To identifiy signs of nonthermal emission, we have also searched Stokes-\textit{V} maps for detections of circularly polarized soources, a telltale sign for gyrosynchrotron radiation. No sources were detected above 5$\sigma$ in a Stokes-$V$ map; the 3$\sigma$ upper limits for the two brightest YSOs are 45~$\mu$Jy, corresponding to upper limits for the circular polarization of LRL 52590 and LRL 49 of 3.5\% and 8.2\%.
Additionally, we are particularly interested in radio counterparts with negative spectral indices, again indicative of nonthermal emission. However, we do not find clear cases of negative spectral indices. While LRL 13 only has a C band counterpart and remains undetected in the X band, the detection is too weak for the upper limit of the spectral index to be significant.

The low number of detections correspond to low detection rates since considerably more YSOs are located in the primary beam areas. Table~\ref{tbl-radio-ic348} lists the different source types located in the primary beam areas of the X-band and C-band observations as well as the respective radio detections. As noted above, the C-band primary beam is larger than the X-band primary beam. All YSO classes have very low detection rates. Overall, only 10\% of the class 0-II YSOs in the C-band primary beam areas were detected.

Our list of detections contains two of the four sources reported by \citet{avi01} in a deeper observation. For the two undetected sources, we can report upper limits. VLA\,1 was reported with an X-band flux density of 0.14~mJy where we find an upper limit of $<0.07$~mJy (5$\sigma$), and for VLA\,4, the reported X-band flux density of 0.59~mJy compares to our upper limit of $<0.08$~mJy (5$\sigma$). These two sources appear to have no known counterparts at other wavelengths and probably are variable extragalactic radio sources. Both VLA\,2 and VLA\,3 also show variability on the time scales of years. VLA\,2 is slightly weaker, but statistically marginally compatible with the previous result while VLA\,3 has become brighter. We summarize this comparison in Table~\ref{tbl-vla-i348-prev}. Among our radio detections, only one source is located in the primary beam area of the \citet{avi01} observations, i.e., source 4, or LRL\,13. However, we only detect the source in C band while \citet{avi01} only discuss X band data.

When imaging the two epochs of this experiment separately, some variability on time scales of days becomes apparent. Only two of the sources seen in the full dataset are detected at $>5\sigma$ in the single epochs and are listed in Table~\ref{tbl-vla-i348-epochs}. Interestingly, we also detect and list two more sources that are detected above the same threshold in only one of the two epochs and remain below the 5$\sigma$ cutoff in the combined data (note that these are not included in Table~\ref{tbl-radio-ic348}). One is LRL 1888, listed by \citet{mue07} as a candidate class III member, and the other is unidentified, though within 1$''$ of a previously listed infrared source ([PSZ2003] J034403.6+320520). Both detections are too weak for reliable determinations of spectral indices.

\begin{deluxetable}{rrrrrrl}
\tablecaption{IC 348: radio detections\label{tbl-vla-i348}}
\tablecolumns{7}
\tablehead{
\colhead{no.} & \colhead{Pos} & \colhead{RA, Dec}   & \colhead{$S_X$}  & \colhead{$S_C$} & \colhead{$\alpha$} & \colhead{ID}\\
\colhead{}    & \colhead{}    & \colhead{(J2000.0)}  & \colhead{(mJy)}  & \colhead{(mJy)} & \colhead{(X-C)}    & \colhead{(lit.)}
}
\startdata
1 & 2 & 3:43:56.816 +32:00:50.06 & $0.087\pm 0.022$    & $0.040\pm0.016$     &                & VLA\,2, HH\,211-mm (0)\\
2 & 2 & 3:43:57.099 +32:03:03.79 & $0.079\pm 0.015$    & $<0.07$ (5$\sigma$) &                & \\ % (\tablenotemark{a})
3 & 2 & 3:43:57.610 +32:01:37.39 & $0.552\pm 0.017$    & $0.542\pm 0.015$    & $0.03\pm 0.11$ & VLA\,3, LRL\,49 (STAR)\\
4 & 2 & 3:43:59.707 +32:01:53.43 & $<0.08$ (5$\sigma$) & $0.087\pm 0.014$    &                & LRL\,13 (THICK,II)\\
5 & 1 & 3:44:12.990 +32:01:35.71 & $0.057\pm 0.018$    & $0.054\pm 0.017$    &                & LRL\,51 (THICK,I)\\
6 & 1 & 3:44:20.377 +32:01:58.60 & $1.312\pm 0.021$    & $1.146\pm 0.019$    & $0.25\pm 0.06$ & LRL\,52590 (I) \\
7 & 1 & 3:44:31.459 +32:00:39.96 & out of pb           & $0.227\pm 0.026$    & --             & no IR counterpart\\
\enddata
\tablecomments{The spectral index is only given when the corresponding error is $\le0.2$.  The second column refers to a detection in either of the two pointings, as defined in the text. The VLA designations are from \citet{avi01} and the LRL numbers and classifications are from \citet{lad06} and \citet{mue07}. The two thick disk sources from \citet{lad06} have different classes according to the spectral index criterion used by \citet{mue07}. For HH 211-mm, see \citet{fro03}.}
\end{deluxetable}

\begin{deluxetable}{llll}
\tablecaption{IC 348: Radio -- detected members\label{tbl-radio-ic348}}
\tablecolumns{4}
\tablehead{
\colhead{} & \colhead{FOV (C)} & \colhead{FOV (X)}& \colhead{det. in X or C}
}
\startdata
Class 0      &  3 &  3 &  1 \\
Class I      & 14 & 14 &  2 \\ 
Class II\tablenotemark{a}     & 22 & 12 &  1 \\
Class III\tablenotemark{b}    & 28 & 13 &  1 \\ 
\enddata
\tablecomments{Based on \citet{lad06} and \citet{mue07}, refined with class 0 sources from \citet{eno09}.}
\tablenotetext{a}{including SED type THICK from \citet{lad06}, apart from sources LRL 51 and LRL 276 which are class~I, see text.}
\tablenotetext{b}{including SED types ANEMIC and STAR from \citet{lad06}}
\end{deluxetable}

\begin{deluxetable}{lrrr}
\tablecaption{IC 348: detections of previously known radio sources\label{tbl-vla-i348-prev}}
\tablecolumns{4}
\tablehead{
\colhead{Src} & \colhead{Position} & \colhead{$S_X$}         & \colhead{$S_X$} \\
\colhead{HH 211--}    & \colhead{(J2000)}  & \colhead{\citep{avi01}} & \colhead{(this work)} 
}
\startdata
VLA 1 & 03 43 47.58 +31 59 41.45 & 0.14~mJy & $<0.07$~mJy ($5\sigma$)\\
VLA 2 & 03 43 56.80 +32 00 50.41 & 0.09~mJy & $0.040\pm0.016$~mJy \\
VLA 3 & 03 43 57.60 +32 01 37.52 & 0.41~mJy & $0.542\pm0.015$~mJy \\
VLA 4 & 03 44 12.10 +31 58 29.32 & 0.59~mJy & $<0.08$~mJy ($5\sigma$) \\
\enddata
\tablecomments{The source positions are from \citet{avi01} but have been converted to J2000.}
\end{deluxetable}

\begin{deluxetable}{lllllr}
\tablecaption{IC 348: radio detections by epoch\label{tbl-vla-i348-epochs}}
\tablecolumns{6}
\tablehead{
\colhead{no.} & \colhead{Pos} & \colhead{ID, RA/Dec}  & \colhead{Epochs 1\&2}     & \colhead{}             & \colhead{}\\
\colhead{}    & \colhead{}    & \colhead{}     & \colhead{$S_X$ (mJy)} & \colhead{$S_C$ (mJy)}  & \colhead{$\alpha$}
}
\startdata
3 & 2 & LRL 49 (STAR)           & Ep 1: $0.829\pm 0.025$    & $0.726\pm 0.022$     & $0.24\pm 0.11$ \\
  &   &                         & Ep 2: $0.364\pm 0.024$    & $0.417\pm 0.019$     &$-0.25\pm 0.20$ \\  
6 & 1 & LRL 52590 (I)           & Ep 1: $1.472\pm 0.033$    & $1.257\pm 0.028$     & $0.29\pm 0.08$ \\
  &   &                         & Ep 2: $1.199\pm 0.030$    & $1.095\pm 0.023$     & $0.17\pm 0.08$ \\
- & 1 & 3:44:19.797 +31:59:18.63& Ep 1: $<0.106$ (5$\sigma$) & $<0.127$ (5$\sigma$) & --            \\
  &   & (LRL 1888, III)         & Ep 2:  $0.175\pm 0.025$    & $0.133\pm 0.022$    &                \\
- & 2 & 3:44:03.652 +32:05:20.67& Ep 1: $<0.080$ (5$\sigma$) & $0.179\pm 0.031$    &              \\
  &   &                         & Ep 2: $<0.097$ (5$\sigma$)& $<0.080$ (5$\sigma$) & --\\
\enddata
\tablecomments{Source numbers refer to Table~\ref{tbl-vla-i348}. The spectral index is only given when the corresponding error is $\le0.2$.}
\end{deluxetable}

%%%%%%%%%%%%%%%%%%%%%%%%%%%%%%%%%%%%%%%%%
%%%%%%%%%%%%%%%%%%%%%%%%%%%%%%%%%%%%%%%%%
%%%%%%%%%%%%%%%%%%%%%%%%%%%%%%%%%%%%%%%%%

\subsubsection{NGC 1333}

We have used the same methods that were used for IC 348 for source detection in NGC 1333. An automatic search above a significance cutoff of 5$\sigma$ was followed by a manual search for lower-significance detections toward previously reported radio sources \citep{rod99} and YSOs \citep{gut08}; the typical map rms is 20~$\mu$Jy. Nine sources were detected by the automatic search while ten more sources were found manually. The resulting source list with 19 sources is shown in Table~\ref{tbl-vla-n1333}. 

At detection significance levels of $3\sigma$, used in the manual search, the number of false detections due to noise peaks is no longer negligible. When using the total number of sources that we searched manually and assuming Gaussian noise, we expect $\sim$1--3 false detections in the X and C bands due to chance alignments of candidate sources and positive image pixels with $>3\sigma$ significance within the area of a synthesized beam. Note that this only holds for detections in just one of the two bands. While Table~\ref{tbl-vla-n1333} also contains sources with single-band detections at levels of $<5\sigma$, these could thus be chance alignments. 

In addition to such alignments with noise peaks, there are also detections of extragalactic sources in the field of view of both clusters. Above the respective 5$\sigma$ detection limits, we can expect about one extragalactic source in each X-band primary beam area and about three extragalactic sources in each C-band primary beam area \citep{win93}. The probability of a chance alignment of a known YSO in the manual search with an extragalactic source is, however, exceedingly low.

Out of 19 sources, ten can be identified with YSOs from \citet{gut08}. One additional source, VLA 2a/b, has a millimeter counterpart has been described by \citet{cha00} and \citet{che09} as a class 0 protostar, but it is an unresolved binary in our data. The source does not appear to have a mid-infrared (\textit{Spitzer}-IRAC) counterpart and was not reported by \citet{gut08}. Two of these YSOs have clearly negative spectral indices, the unresolved source VLA 2a/b and [GMM2008] 97. Source 10, while not selected as a \textit{Spitzer} YSO, has been described as an M2.7 dwarf star by \citet{wil04}. Stokes-$V$ maps show that none of the NGC 1333 sources reported here have significant circular polarization.

Previously, \citet{rod97} reported results from a comparable VLA observation of the region, resulting in the detection of sources VLA 1--4, all of which we can confirm, although the flux densities appear to have changed (Table~\ref{tbl-vla-n1333-prev}). Additionally, these detections can be compared to the deep radio study of the region by \citet{rod99}. Their map is twice as deep as ours, with an rms noise of 10~$\mu$Jy, and they detect 35 sources above a significance of 5~$\sigma$. These sources include 16 X-band detections above 0.10~mJy, corresponding to the 5$\sigma$ cutoff used in our data. All but one of these sources are in the primary beam area (FWHM) of our southern pointing, but many remain undetected. However, \citet{rod99} use a concatenated dataset consisting of several VLA observations in different configurations, requiring the use a common subset in the {\it uv} plane so that the beam size is $\sim5''$, emphasizing more extended structures than our setup. We have searched all 20 X-band detections reported by \citet{rod99} for counterparts in our X-band data, and we find five additional radio sources that did not make our SNR cutoff; see Table~\ref{tbl-vla-n1333-prev} where we also compare the flux densities to the values given by \citet{rod97,rod99}. Only one source has a flux density that is statistically compatible with the previously reported value (within 1$\sigma$). 

Finally, we report the radio detections of the different YSO evolutionary classes in Table~\ref{tbl-radio-n1333}. The table lists the YSOs located in the primary beam areas and the detection counts. The resulting detection fractions are again low, if marginally higher than in IC~348. Across classes 0--II, 13\% of the YSOs located in the C-band primary beam areas have been detected in either radio band.

We have again checked for source variability by separately imaging the two epochs. Only two sources are detected above a significance of 5$\sigma$ in the two individual epochs, and they are listed in Table~\ref{tbl-vla-n1333-epochs}. Both sources, VLA 2a/b and [GMM2008] 97 show some variability and have consistently negative spectral indices.

\begin{deluxetable}{lrrrrrl}
\tablecaption{NGC 1333: radio detections\label{tbl-vla-n1333}}
\tablecolumns{7}
\tablehead{
\colhead{no.} & \colhead{N/S} & \colhead{RA, Dec}   & \colhead{$S_X$}  & \colhead{$S_C$} & \colhead{$\alpha$}  & \colhead{ID}\\
\colhead{}    & \colhead{}    & \colhead{(J2000.0)}  & \colhead{(mJy)}  & \colhead{(mJy)} & \colhead{(X-C)}    & \colhead{(lit.)}
}
\startdata
 8  & S & 3:28:57.382 +31:14:15.79 & $<0.11$ (5$\sigma$) & $0.164\pm 0.025$    &                 & [G08] 3 (I*) \\
 9  & S & 3:28:57.650 +31:15:31.33 & $0.307\pm 0.026$    & $0.483\pm 0.023$    & $-0.83\pm 0.24$ & VLA 1, extragal.  \\
10  & S & 3:29:00.253 +31:13:39.19 & $0.170\pm0.037$     & $<0.14$ (5$\sigma$) &                 & VLA\,13, (M2.7\tablenotemark{a})\\
11  & S & 3:29:01.964 +31:15:38.04 & $1.073\pm 0.021$    & $1.237\pm 0.021$    & $-0.26\pm 0.07$ & VLA 2a/b, MMS 3 (0)\\
12  & S & 3:29:01.970 +31:15:36.22 & $0.093\pm 0.021$    & $0.185\pm 0.022$    &                 & \\
13  & S & 3:29:03.082 +31:15:51.66 & $0.078\pm0.020$     & $<0.12$ (5$\sigma$) &                 & VLA 17\\
14  & S & 3:29:03.372 +31:16:01.60 & $0.100\pm0.021$     & $0.122\pm0.021$     &                 & VLA 3\\
15  & S & 3:29:03.756 +31:16:03.88 & $0.164\pm 0.020$    & $0.107\pm 0.025$    &                 & VLA\,4,\,[G08]\,29\,(I)  \\
16  & S & 3:29:05.742 +31:16:39.71 & $0.077\pm0.021$     & $<0.11$ (5$\sigma$) &                 & VLA\,22,\,[G08]\,84\,(II)\\
17  & N & 3:29:07.772 +31:21:57.03 & $0.130\pm0.019$     & $<0.12$ (5$\sigma$) &                 & [G08]  31 (I) \\
18  & N & 3:29:10.373 +31:21:58.94 & $0.442\pm 0.020$    & $0.573\pm 0.022$    & $-0.48\pm 0.15$ & [G08] 97 (II) \\
19  & S & 3:29:10.420 +31:13:32.09 & $0.179\pm0.041$     & $<0.13$ (5$\sigma$) &                 & VLA\,25,\,[G08]\,6\,(0\tablenotemark{b})\\
20  & S & 3:29:11.250 +31:18:30.99 & $<0.26$ (5$\sigma$) & $0.171\pm0.028$     &                 & [G08]   7 (0\tablenotemark{b})\\
21  & N & 3:29:17.678 +31:22:45.49 & $0.055\pm0.018$     & $0.077\pm0.022$     &                 & [G08] 111 (II)\\
22  & S & 3:29:20.643 +31:15:49.59 & out of pb           & $0.816\pm 0.029$    & --              & VLA 32, extragal.\\
23  & S & 3:29:20.908 +31:15:49.36 & out of pb           & $0.169\pm 0.029$    & --              & \\
24  & N & 3:29:21.328 +31:23:46.20 & $<0.14$ (5$\sigma$) & $0.089\pm0.024$     &                 & [G08] 117 (II)\\
25  & S & 3:29:22.266 +31:13:54.40 & $<0.77$ (5$\sigma$) & $0.178\pm0.036$     &                 & [G08]  38 (I) \\
26  & N & 3:29:30.923 +31:22:11.82 & out of pb           & $0.479\pm 0.033$    & --              &  \\
\enddata
\tablecomments{Running number continued from Table~\ref{tbl-vla-i348}. The second column refers to a detection in either the northern (N) or the southern (S) pointing. The spectral index is only given when the corresponding error is $\le0.3$. [GMM2008] in ID column abbreviated to [G08]; the VLA designations are from \citet{rod97,rod99}. For the evolutionary state of source 11, see \citet{cha00,che09}.}
\tablenotetext{a}{from \citet{wil04}.}
\tablenotetext{b}{from \citet{eno09}.}
\end{deluxetable}

\begin{deluxetable}{lrrrrrr}
\tablecaption{NGC 1333: detections of previously known X-band radio sources\label{tbl-vla-n1333-prev}}
\tablecolumns{6}
\tablehead{
\colhead{Src} & \colhead{Position} & \colhead{$S_X$(total)}        & \colhead{$S_X$(peak)} & \colhead{$S_C$(total)}	   & \colhead{$S_C$(peak)} \\
\colhead{HH 7-11--} & \colhead{(J2000)}  & \colhead{(mJy)\tablenotemark{a}} & \colhead{(mJy, this work)} & \colhead{(mJy)\tablenotemark{a}} & \colhead{(mJy, this work)}
}
\startdata
VLA 1    & 03:28:57.65 +31:15:31.5 & 0.52 & $0.307\pm0.026$ & 1.05  & $0.483\pm0.023$	  \\
VLA 2a   & 03:29:01.95 +31:15:38.3 & 1.43 & $1.073\pm0.021$ & 1.36  & $1.237\pm0.021$	  \\
VLA 3    & 03:29:03.37 +31:16:01.9 & 0.27 & $0.100\pm0.022$ & 0.12  & $0.122\pm0.021$	  \\
VLA 4    & 03:29:03.75 +31:16:04.1 & 0.32 & $0.164\pm0.020$ & 0.21  & $0.107\pm0.025$	  \\
VLA 13   & 03:29:00.29 +31:13:38.6 & 0.13 & $0.170\pm0.037$ &$<0.06$& $<0.14$ (5$\sigma$) \\
VLA 17   & 03:29:03.08 +31:15:52.5 & 0.08 & $0.078\pm0.020$ &$<0.04$& $<0.12$ (5$\sigma$) \\
VLA 22   & 03:29:05.74 +31:16:39.5 & 0.17 & $0.073\pm0.020$ & 0.17  & $<0.11$ (5$\sigma$) \\
VLA 25   & 03:29:10.46 +31:13:32.1 & 0.49 & $0.179\pm0.041$ & 0.36  & $<0.13$ (5$\sigma$) \\
VLA 32   & 03:29:20.67 +31:15:49.6 & 2.55 & out of pb	    & 3.64  & $0.816\pm0.029$	  \\
\enddata
\tablecomments{The source positions are from \citet{rod97} and \citet{rod99} but have been converted to J2000. Note that the flux densities from \citep{rod99} are total fluxes while we quote peak flux densities. Flux densities have been corrected for primary beam attenuation.}
\tablenotetext{a}{from \citet{rod97,rod99}, flux density errors are listed there}
\end{deluxetable}

\begin{deluxetable}{llll}
\tablecaption{NGC 1333: Radio -- detected members\label{tbl-radio-n1333}}
\tablecolumns{4}
\tablehead{
\colhead{} & \colhead{FOV (C)} & \colhead{FOV (X)}& \colhead{det.}
}
\startdata
Class 0     &  9 &  5 &  3 \\
Class I     & 20 &  8 &  4 \\
Class II    & 57 & 21 &  4 \\
Class II/III&  2 &  1 &  0 \\
\enddata
\tablecomments{YSOs from \citet{gut08}, refined with class 0 sources from \citet{eno09}.}
\end{deluxetable}

\begin{deluxetable}{llllll}
\tablecaption{NGC 1333: radio detections by epoch\label{tbl-vla-n1333-epochs}}
\tablecolumns{6}
\tablehead{
\colhead{no.} & \colhead{Pos} &\colhead{ID}    & \colhead{$S_X$ (mJy)}  & \colhead{$S_C$ (mJy)}             & \colhead{$\alpha$}        \\
}
\startdata
11 & S & VLA 2a/b               & Ep 1: $1.216\pm 0.034$     & $1.228\pm 0.030$     & $-0.02\pm 0.10$\\
   &   &                        & Ep 2: $1.021\pm 0.026$     & $1.236\pm 0.026$     & $-0.35\pm 0.09$\\
18 & N & [GMM2008] 97 (II)      & Ep 1: $0.613\pm 0.034$     & $0.699\pm 0.030$     & $-0.24\pm 0.18$\\
   &   &                        & Ep 2: $0.336\pm 0.024$     & $0.450\pm 0.026$     & $-0.54\pm 0.24$\\
\enddata
\end{deluxetable}

\subsubsection{Summary}

Between the two clusters IC 348 and NGC 1333, we report the detections of radio counterparts toward 16 YSOs, 11 of these are located in NGC 1333. None of these could be unambiguously identified as nonthermal radio sources due to polarization. Using simultaneously measured negative radio spectral indices and variability instead as proxies for nonthermal emission yields two nonthermal radio counterparts in NGC 1333 (VLA 2a/b and [GMM2008] 97). Vice versa, only a single source, the class I source LRL 52590, clearly appears to be a thermal source when judged by its radio spectral index of $\alpha=0.25\pm 0.06$.

%%%%%%%%%%%%%%%%%%%%%%%%%%%%%%%%%%%%%%%%%%%%%%%%%%%%%%%%%%%%%%%%%%%%%%%%%%%%%%%%%%%%%%

\subsection{X-ray data}

The \textit{Chandra} X-ray observations cover a considerably larger area than the VLA observations. In the following, we mainly focus on the \textit{Chandra} X-ray data covering the VLA primary beam areas.

\subsubsection{IC 348}
In order to reduce and process the X-ray data, we have used the ANCHORS pipeline. Briefly, the two observations are combined for the purpose of detecting point sources. Two iterations with the ciao wavelet-based tool wavdetect were performed. The first pass took a square region 15$'$ on each side and centered at the aimpoint at full resolution. The second pass examined a larger (25$'\times$25$'$) area including all six ACIS chips with pixels binned by two. To measure source parameters, the two observations were analyzed separately. For the extraction regions of each source, the parameters of the ellipse-shaped point-spread function are interpolated from a lookup table \citep{all04} for a 95\% encircled energy radius. The total counts are then scaled to 100\%.

In the combined dataset, 149 sources were identified. In the two single datasets, 134 sources and 109 X-ray sources were found in datasets 8933 and 8944, respectively. The field of view of the \textsl{Chandra} observations covers 60 YSOs listed in \citet[][Table 2]{lad06}, and also 30 YSOs listed in \citet[][Tables 1, 2, and 11]{mue07}. Out of this total of 90 YSOs, including 5 candidate class III sources, 33 sources have X-ray counterparts within a search radius of $2''$. Table~\ref{tbl-xray} lists the statistics of X-ray detections as a functions of object types, concatenated into totals for the different YSO classes. The table also contains information on the detection statistics for the nominal primary beam areas of the VLA observations.

The X-ray detection fraction for the YSOs is $\sim$1/3 in object classes I--III. None of the class 0 sources were detected in X-rays. Table~\ref{tbl-src} lists the X-ray sources with counterparts in the YSO samples. Listed are the positional difference of the X-ray detections compared to the sample, IDs, X-ray counts in both observations and information on variability, spectral and SED types. Apart from the 33 above-mentioned sources, the table lists two additional stellar sources listed by \citet{lad06} without an evolutionary class. LRL 1916, a class I source, shows the most spectacular variability between the two observations. The source is detected at 75 counts in the first observation, then drops to a nominal single count in the second observation.

\begin{deluxetable}{lrrrr}
\tablecaption{IC 348: X-ray -- detected YSOs\label{tbl-xray}}
\tablecolumns{5}
\tablehead{
\colhead{}   & \colhead{in FOV-CXO} & \colhead{w/X-ray} & \colhead{also in FOV-VLA(C)} & \colhead{w/X-ray}
}
\startdata
Class 0      &  3 &	  0 &		 3 &	   0 \\
Class I      & 15 &	  7 &		14 &	   5 \\
Class II     & 29 &	 10 &		21 &	   5 \\
Class III    & 37 &	 16 &		24 &	  10 \\
%               ?        OK             
\enddata
\end{deluxetable}

\begin{deluxetable}{rrrrrrrrl}
\small
\tablecaption{IC 348: X-ray--detected YSOs\label{tbl-src}}
\tablecolumns{9}
\tablehead{
\colhead{LRL} & \colhead{RA}        & \colhead{Dec}       & \colhead{Offset}  & \colhead{X-ray cnts.} & \colhead{X-ray cnts.} & \colhead{cnt.} & \colhead{SpT} & \colhead{SED type}\\
\colhead{}    & \colhead{(\textsl{Chandra})} & \colhead{(J2000.0)} & \colhead{($''$)} & \colhead{(8933,raw)}      & \colhead{(8944,raw)}      & \colhead{ratio}    & \colhead{}    & \colhead{(IRAC)}
}
\startdata
1840  & 03:43:19.91 & +32:02:43.13 & 1.75 & 33  & 24  & 1.4  &        & III       \\ %N
273   & 03:43:52.01 & +32:03:40.45 & 1.11 & 20  & 16  & 1.3  &        & III       \\ %N
124   & 03:43:54.65 & +32:00:29.96 & 0.24 & 27  & 42  & 1.6  & M4.25  & STAR	  \\
26    & 03:43:56.06 & +32:02:13.35 & 0.37 & 19  & 12  & 1.6  & K7     & THICK	  \\
49    & 03:43:57.62 & +32:01:37.47 & 0.33 & 56  & 57  & 1.0  & M0.5   & STAR	  \\
13    & 03:43:59.67 & +32:01:54.10 & 0.38 & 182 & 107 & 1.7  & M0.5   & THICK	  \\
160   & 03:44:02.62 & +32:01:35.06 & 0.37 & 29  & 21  & 1.4  & M4.75  & STAR	  \\
54460 & 03:44:02.66 & +32:01:59.35 & 0.54 & 4	& 11  & 2.8  &        & I	  \\
22021 & 03:44:03.64 & +32:02:33.25 & 0.31 & 23  & 17  & 1.4  & M5     & 	  \\
1916  & 03:44:05.79 & +32:00:28.55 & 0.14 & 75  & 1   & 75.0 &        & I	  \\
214   & 03:44:07.54 & +32:04:08.65 & 0.39 & 10  & 17  & 1.7  & M4.75  & THICK	  \\
276   & 03:44:09.22 & +32:02:38.14 & 0.36 & 15  & 13  & 1.2  & M0     & THICK\tablenotemark{a}	  \\ %class I
173   & 03:44:10.23 & +32:04:05.53 & 1.62 & 6   & 18  & 3.0  & M5.75  & THICK     \\ %N
105   & 03:44:11.30 & +32:06:12.31 & 0.57 & 112 & 83  & 1.3  & M0     & STAR	  \\
51    & 03:44:13.00 & +32:01:35.51 & 0.43 & 14  & 31  & 2.2  &        & THICK\tablenotemark{a}	  \\ %class I
54519 & 03:44:13.78 & +31:55:34.85 & 0.47 & 92  & 86  & 1.1  &        & III	  \\
31    & 03:44:18.21 & +32:04:57.26 & 0.65 & 219 & 136 & 1.6  & G1     & ANEMIC    \\
99    & 03:44:19.33 & +32:07:34.11 & 1.12 & 105 & 171 & 1.6  & M3.75  & THICK     \\ %N
1888  & 03:44:19.78 & +31:59:19.05 & 0.05 & 46  & 159 & 3.5  &        & III	  \\
210   & 03:44:20.01 & +32:06:45.49 & 0.15 & 45  & 47  & 1.0  & M3.5   & STAR	  \\
52590 & 03:44:20.43 & +32:01:57.93 & 0.78 & 50  & 75  & 1.5  &        & I	  \\
119   & 03:44:21.29 & +32:05:02.66 & 0.49 & 91  & 153 & 1.7  & M2.5   & STAR	  \\
1889  & 03:44:21.35 & +31:59:32.80 & 0.10 & 15  & 13  & 1.2  &        & I	  \\
125   & 03:44:21.62 & +32:06:25.71 & 1.07 & 70  & 171 & 2.4  & M2.75  & STAR      \\ %N
61    & 03:44:22.28 & +32:05:43.23 & 0.54 & 54  & 80  & 1.5  & K8     & THICK	  \\
72    & 03:44:22.58 & +32:01:53.75 & 0.07 & 81  & 130 & 1.6  & M2.5   & ANEMIC    \\
30190 & 03:44:23.65 & +32:06:47.67 & 0.87 & 158 & 158 & 1.0  & M2.5   & 	  \\
123   & 03:44:24.60 & +32:03:56.97 & 0.45 & 119 & 231 & 1.9  & M1     & STAR	  \\
97    & 03:44:25.58 & +32:06:17.14 & 0.33 & 39  & 699 & 17.9 & M2.25  & THICK	  \\
5     & 03:44:26.05 & +32:04:30.18 & 0.37 & 756 & 527 & 1.4  & G8     & THICK	  \\
62    & 03:44:26.68 & +32:03:58.52 & 0.70 & 103 & 96  & 1.1  & M4.75  & STAR	  \\
69    & 03:44:27.05 & +32:04:43.58 & 0.43 & 620 & 471 & 1.3  & M1     & STAR	  \\
68    & 03:44:28.49 & +31:59:54.52 & 0.49 & 10  & 18  & 1.8  & M3.5   & THICK	  \\
55    & 03:44:31.40 & +32:00:14.16 & 0.79 & 60  & 153 & 2.6  & M0.5   & THICK	  \\
52648 & 03:44:34.44 & +31:58:00.46 & 1.05 & 50  & 41  & 1.2  &        & I         \\ %N
\enddata
\tablecomments{Spectral types and classification information are from \citet{lad06} and \citet{mue07}.}
\tablenotetext{a}{class I, see text}
\end{deluxetable}

\subsubsection{NGC 1333}
The data reduction of the NGC~1333 X-ray data is described by \citet{win10}. These authors find X-ray counterparts to 54 previously known YSOs and identify 41 additional X-ray sources with infrared counterparts as diskless class~III YSOs. The field of view of the \textsl{Chandra} observations covers 123 out of 137 candidate YSOs listed by \citet{gut08}. As discussed by \citet{win10}, roughly a third of the class~I and flat-spectrum protostars and half of the class~II sources are detected in X-rays. Again, none of the class 0 sources have X-ray counterparts.

\begin{deluxetable}{lrrrr}
\tablecaption{NGC 1333: X-ray -- detected class 0--II YSOs in NGC~1333 \label{tbl-xray-n1333}}
\tablecolumns{5}
\tablehead{
\colhead{}   & \colhead{in FOV-CXO} & \colhead{w/X-ray} & \colhead{also in FOV-VLA(C)} & \colhead{w/X-ray} % gut08 correalted with Win10
}
\startdata
Class 0   &     7 &       0 &             7 &       0 \\
Class I   &    26 &       6 &            21 &       5 \\ %
Class II  &    83 &      44 &            74 &      38 \\ %
\enddata
\tablecomments{X-ray data from \citet{win10}.}
\end{deluxetable}

%%%%%%%%%%%%%%%%%%%%%%%%%%%%%%%%%%%%%%%%%%%%%%%%%%%%%%%%%%%%%%%%%%%%%%%%%%%%%%%%%%%%%%%%%%%%%%%%%%%%%%%%%%%%%%%%%%%%%%%%%%

\section{Discussion}

\subsection{Combined X-ray and radio properties}

The combination of the X-ray and radio data has been performed with a search radius of 1$''$. Using larger search radii up to 4$''$ did not increase the number of sources detected in both bands. In total, eight YSOs have been detected in both bands, four in each cluster. The X-ray fit results from ANCHORS, assuming an absorbed single-temperature APEC plasma, as well as X-ray and radio luminosities for these sources are shown in Table~\ref{tbl-radio-xray}. The X-ray and radio source positions coincide within the errors. 

While the VLA and \textit{Chandra} have similar angular resolution, the existence of multiple sources within a scale of about 1$''$ cannot be constrained from such data. Multiplicity could be due to both actual multiple stellar systems and a single object with distinct jet components. As a result, a perceived mismatch of X-ray and radio properties could be due to an unresolved multiple source.

Using the X-ray spectral parameters to predict the level of optically thin thermal free-free radio emission returns predicted flux densities that are up to 10,000 times less than what is observed. Other than these sources, several more are detected in just one of the two bands, but the overwhelming majority of \textit{Spitzer}-selected YSOs are neither detected in X-rays, nor in centimetric radio emission. 

In the following, we will separately assess the radio and X-ray detection fractions for class I and class II YSOs in the two clusters. For the sake of this discussion, we will leave out the class III sources since they are less completely catalogued in both clusters. In order to obtain a comparison between these detection fractions, we only take into account sources in the VLA primary beam (FWHM) areas. As shown in Table~\ref{tbl-xr-comb}, separately for IC\,348 and NGC\,1333, the detection fractions are very diverse. Note that some of the fractions are based on very low numbers and thus are quite uncertain. The radio detection fractions decline with evolutionary stage, similarly in both clusters, but no such trend is apparent in the X-ray detection fractions. The X-ray detection fractions of class I and II YSOs are up to ten times higher than the radio detection fractions, but there is also the curious case of class I protostars in NGC 1333 for which the radio and X-ray detection fractions are similar, but nevertheless due to almost completely distinct sets of sources. Only a minor fraction of YSOs is detected in both bands. The present results thus do not require a tight connection between the observed X-ray and radio emission. 

Compared to the range expected by the GB relation, the X-ray data are much more sensitive than the radio data, placing any radio detection with an X-ray upper limit off the G\"udel-Benz relation by several orders of magnitude. For these data points, it is important to keep in mind that the observations in the two bands were carried out simultaneously so that variability cannot explain this discrepancy.  While some of the YSOs with radio counterparts and X-ray upper limits are close to the radio detection limit and may be chance alignments (see above), there are also some clearly detected radio sources among the YSOs that do not have X-ray counterparts. The most striking example is the class 0 protostar source 11 in NGC 1333 with a radio flux density corresponding to a luminosity of $L_R=7.0\times10^{16}$\,erg\,s$^{-1}$ but no X-ray detection. While here, any X-ray emission is likely extincted, the fact that this source is entirely incompatible with the GB relation may also point to a different radio emission mechanism, even though source 11 is among our candidate nonthermal sources due to its negative spectral index.

Since this comparison is limited by the sensitivity of the radio data presented here, it is too early to draw conclusions concerning the radio populations of the two clusters. However, we can compare our result to previously published observations. Applying the 5$\sigma$ sensitivity limit of the present study to the results on CrA, as reported in the deep radio observations reported in \citet{for06} and scaling to the distance of Perseus, shows that about half of the radio sources there would still be detected. While it thus seems likely that a genuine lack of radio activity plays a role in the low detection rate, the numbers involved are too low to warrant a firm conclusion.

\begin{deluxetable}{lrrrrrr}
\tablecaption{YSOs detected in both the X-ray and the radio range\label{tbl-radio-xray}}
\tablecolumns{7}
\tablehead{
\colhead{Source} & \colhead{Class} &  \colhead{raw cnts.}   & \colhead{$n({\rm H})$} & \colhead{$kT$}& \colhead{$L_X$}& \colhead{$L_R$(X-band\tablenotemark{a})}\\
\colhead{} & \colhead{} & \colhead{(total)}  & \colhead{(10$^{22}$ cm$^{-2}$)} & \colhead{(keV)}& \colhead{erg~s$^{-1}$}& \colhead{erg~s$^{-1}$~Hz$^{-1}$}\\
}
\startdata
LRL 51          & I   &   45 & --            & --            & --                 & $8.6\times10^{15}$\\
{[}GMM2008]  31 & I   &   56 & --            & --            & --                 & $4.3\times10^{15}$\\
LRL 52590       & I   &  125 & $1.87\pm0.30$ & $9.21\pm2.10$ & $4.4\times10^{29}$ & $1.1\times10^{17}$\\
LRL 13          & II  &  289 & $1.42\pm0.25$ & $2.20\pm0.47$ & $8.4\times10^{29}$ & $6.5\times10^{15}$\\
{[}GMM2008]  84 & II  &  455 & $1.36\pm0.15$ & $3.40\pm0.56$ & $7.4\times10^{29}$ & $5.1\times10^{15}$\\
{[}GMM2008]  97 & II  & 3796 & $1.03\pm0.04$ & $1.97\pm0.07$ & $5.7\times10^{30}$ & $3.1\times10^{16}$\\
{[}GMM2008] 111 & II  & 6397 & $0.28\pm0.01$ & $2.46\pm0.08$ & $5.0\times10^{30}$ & $3.6\times10^{15}$\\
LRL 49          & III &  113 & $1.02\pm0.24$ & $2.44\pm0.72$ & $2.9\times10^{29}$ & $4.5\times10^{16}$\\
\enddata
\tablecomments{X-ray information is from ANCHORS for an energy range of 0.3-8.0 keV, spectral fits are only shown for sources with more than 100 counts.}
\tablenotetext{a}{C band for LRL 13}
\end{deluxetable}

\begin{deluxetable}{rlllll}
\tablecaption{Combined Radio-X-ray detection statistics (IC~348+NGC~1333)\label{tbl-xr-comb}}
\tablecolumns{6}
\tablehead{
   & Cluster & FOV (C) & X-ray & radio\tablenotemark{a} & radio(X or C) \& X-ray 
}
\startdata
%IC348+NGC1333
Class 0  & IC 348    &  3 &  0  (0\%) & 1 (33\%) & 0  (0\%)\\
Class I  & IC 348    & 14 &  5 (36\%) & 2 (14\%) & 2 (14\%)\\
Class II & IC 348    & 21 &  5 (24\%) & 1  (5\%) & 1  (5\%)\\
Class 0  & NGC 1333  &  8 &  0  (0\%) & 3 (38\%) & 0  (0\%)\\
Class I  & NGC 1333  & 20 &  5 (25\%) & 4 (20\%) & 1  (5\%)\\
Class II & NGC 1333  & 66 & 38 (58\%) & 4  (6\%) & 3  (5\%)\\
%XXX updated feb 2
\enddata
\tablenotetext{a}{In C-band primary beam area.}
\end{deluxetable}

%\begin{deluxetable}{rllll}
%\tablecaption{Combined Radio-X-ray detection fractions (IC~348/NGC~1333)\label{tbl-xr-comb2}}
%\tablecolumns{5}
%\tablehead{
%      & Cluster & X-ray     & radio & X-ray \& radio
%}
%\startdata
%Class 0  & IC 348      &  0\% & 33\% &  0\% \\
%Class I  & IC 348      & 36\% & 14\% & 14\% \\
%Class II & IC 348      & 24\% &  5\% &  5\% \\
%Class 0  & NGC 1333    &  0\% & 38\% &  0\% \\
%Class I  & NGC 1333    & 25\% & 20\% &  5\% \\
%Class II & NGC 1333    & 58\% &  6\% &  5\% \\
%%XXX updated feb 2
%\enddata
%\end{deluxetable}

\begin{figure*}
\centering
\includegraphics[width=\linewidth]{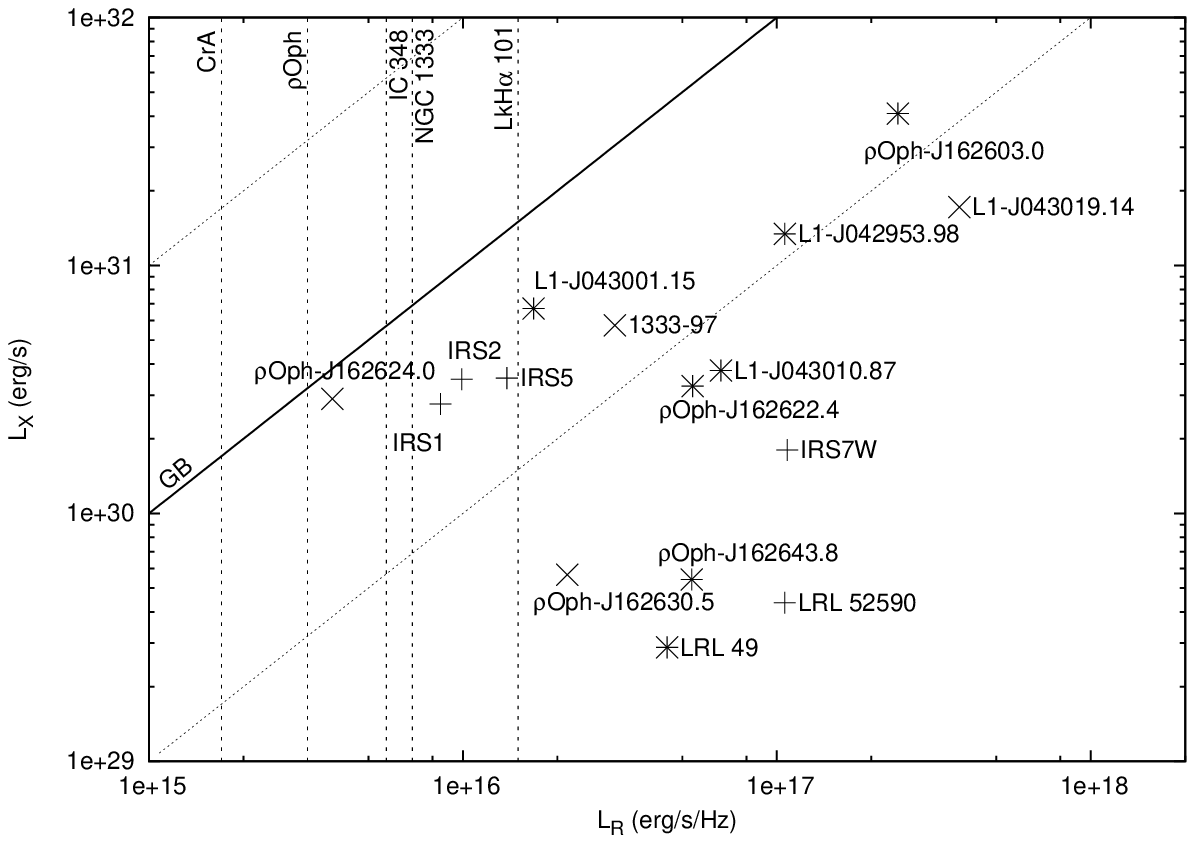}
\caption{Plot of the X-ray and radio luminosities of all YSOs that have been simultaneously detected in both the X-ray and radio ranges by \textsl{Chandra} and the VLA, excluding sources with a radio detection significance of $<5\sigma$ and less than 100 counts in X-rays. The symbols `+', `$\times$', and `\textasteriskcentered' indicate YSO classes I, II, and III, respectively. Apart from sources reported in this paper, and sources in $\rho$ Oph from \citet{gag04}, scaled to a distance of 130~pc, IRS sources are in CrA \citep{for07}, and L1 stands for LkH$\alpha$\,101 \citep{ost09}. The central line marked ``GB'' indicates the G\"udel-Benz relation with its upper and lower bounds, as reported by  \citet{gue02}, shown as dashed lines. The vertical lines indicate the 5$\sigma$ sensitivities of the different observations, as converted into radio luminosities at the respective distances. The limits are strictly for the simultaneous observations only (some regions have deeper radio and/or X-ray data from separate observations). The X-ray sensitivity limits are lower than the X-ray luminosity range shown (see Table~\ref{tbl-cxo-vla}).\label{n1333_ic348_xr_plot2}}
\end{figure*}

Leaving aside the issue of non-detections in one of the two bands, we can compare the YSOs that were detected in both bands with corresponding sources from previous experiments. In spite of the fact that we do not know the dominating emission mechanism for most sources, it is also interesting to compare these YSOs to the empirical GB relation. Figure~\ref{n1333_ic348_xr_plot2} shows all such sources from IC 348, NGC 1333, $\rho$ Oph, CrA, and LkH$\alpha$\,101 \citep{gag04,for07,ost09}, again relative to the GB relation. Also, we conservatively exclude sources with radio detection significance levels of $<5\sigma$ or less than 100 X-ray counts. Note that since the radio luminosities do not depend on a spectral fit, they are dominated by the measurement error. The X-ray luminosities, in contrast, can have uncertainties of order unity. Two things become apparent in this plot: 1) Compared to the GB relation, the YSOs appear to have higher radio luminosities for a given X-ray luminosity, and 2) there are no clear trends with evolutionary stage. About half of the sources are compatible with the GB relation even though there are sources that lie considerably below the GB relation, outside of its approximate bounds of plus or minus one order of magnitude. These sources are of all three evolutionary classes. Since all of these data points are due to simultaneous observations, variability plays only a minor role. 
Also, there are no clear trends concerning the type of emission and compatibility with the GB relation. For example, the thermal source LRL\,52590 is the source that is least compatible with the GB relation in Figure~\ref{n1333_ic348_xr_plot2}. However, also note the opposite example of source 11 above. These differences may indicate different radio emission mechanisms.

In order to judge the significance of this result for YSOs in general it is important to keep in mind that this plot contains only a minor fraction of the YSOs that have been covered in the simultaneous X-ray and radio experiments. While many of the sources plotted appear radio bright, many other sources remain undetected. The detections are due to sources with extreme radio and X-ray luminosities, even though they do not represent all sources in this luminosity range. To illustrate this point, we can use the GB relation to determine the X-ray luminosity that corresponds to the approximate 5$\sigma$ radio detection limit for IC~348 and NGC~1333, $L_X\sim6\times10^{30}$~erg\,s$^{-1}$. This X-ray luminosity limit would include the X-ray brightest $\sim25$\% G and K pre-main sequence stars in Orion \citep{pre05}. None of the X-ray detected YSOs in NGC~1333 and IC 348 actually reach this luminosity. It is, however, noteworthy that six infrared-identified YSOs in NGC 1333 (from \citealp{gut08}) have X-ray luminosities up to one order of magnitude below this number and lie within the VLA primary beam areas, but they do not have radio counterparts. A similar trend can be seen in the data presented by \citet{gag04}. There, in $\rho$ Oph, 26 X-ray sources are more luminous than the expectation for the lower limit of the GB relation at the 5$\sigma$ radio detection limit. Of these X-ray sources, only 5 have radio counterparts. Given the radio sensitivity limits of the various studies, as shown in Figure~\ref{n1333_ic348_xr_plot2}, sources that are undetected in the radio range could, however, well be compatible with the GB relation at lower luminosities.

A shift toward higher radio luminosities for given X-ray luminosities had already been noted by \citet{gue93} who found that the most luminous sources were slightly offset toward higher radio luminosities, yet still within the quoted limits of the relation. \citet{gue97} suggest that not only a high X-ray luminosity, but rather the presence of a hot plasma component is conditional for the detectability of a late-type star as a strong, nonthermal radio source. However, since we also find indications for YSOs with radio but no X-ray counterparts, this may not directly apply to YSOs. They also find that while X-ray emission saturates that does not seem to be the case for the non-thermal radio emission. Consequently, if all sources discussed here were X-ray--saturated, that could also explain an offset from the GB relation. However, it seems unlikely that this would explain the offsets shown in Figure~\ref{n1333_ic348_xr_plot2} where the YSOs that are furthest off the GB relation have the lowest X-ray luminosities.

One important caveat in these comparisons to the GB relation is the nature of the radio emission for each single detection since the correlation is only valid for nonthermal emission. Among the sources reported in this paper, no source can conclusively be proven to be nonthermal due to the relatively low signal-to-noise ratios: No polarization was detected, and the error bars on the radio spectral indices also do not allow us to use them as definitive criteria. Among the class~I sources in the literature, CrA-IRS~5 is known to show non-thermal radio emission \citep{fei98}. This source is compatible with the GB relation \citep{for06,for07}. Potentially, the outlier sources have additional or alternative thermal radio components that would shift them toward higher radio luminosities. This thermal component would, however, in some cases have to account for several orders of magnitude in luminosity. A third and again completely different possibility involves coherent emission processes which, while present on the Sun \citep{ben05,ben06}, have not yet been identified in YSOs. 

While also the thermal X-ray emission of YSOs is known to be due to at least two different sources (coronal magnetic activity and accretion), this results primarily in different temperature components in the X-ray emission, generally of similar luminosities. For our comparison of X-ray and radio luminosities, the interpretation of X-ray luminosities should thus be more straightforward.

\section{Summary and Conclusions}

We present simultaneous X-ray and radio observations of IC~348 and NGC~1333. Only three other clusters have previously been observed in this way. In summary, our main conclusions are as follows:

\begin{itemize}
\item The radio detection fraction for class I and class II YSOs is generally much lower than the X-ray detection fraction, particularly for the class~II sources. The radio detections are not solely a subset of the X-ray detected sources. Very few YSOs ($\sim5$\%) are detected in both bands. The percentage of such X-ray--radio detections is marginally higher for class I protostars than for class II sources.
\item We find a few YSOs that are detected in only one band (X-ray or radio), even though they should be easily detectable in the other band if the GB relation is applicable to these sources. Extinction helps to explain X-ray non-detections, but whether radio non-detections of X-ray sources are due to insufficient sensitivity remains unclear.
\item The currently known sample of YSOs that have been simultaneously detected in both the X-ray and radio bands (across several clusters) shows that the corresponding luminosities of only about half of these sources are within one order of magnitude of the GB relation while the other half is shifted toward higher radio luminosities for a given X-ray luminosity.
\item Compared to early results on the GB relation almost 20 years ago, simultaneous X-ray and radio data are now available for several source populations, minimizing the effect of variability. Also, the X-ray properties of YSOs in particular are much better defined while there has been comparably little progress on their centimetric radio properties.
\item A key uncertainty in these comparisons is the nature of the radio emission since these relatively weak radio sources cannot unambiguously be identified as nonthermal radio sources. Also the radio luminosity function for these sources is basically unknown. Radio sensitivity strongly limits the size of the dataset which can be compared to the GB relation. With the advent of the Expanded Very Large Array, the number of YSOs with defined radio properties can be expected to increase significantly. With such observations, it will be possible to look for fainter radio sources and to determine the radio luminosity function for entire YSO populations.
\end{itemize}

\acknowledgments{Support for this work was provided by the National Aeronautics and Space Administration through Chandra Award Numbers GO6-7015X and GO8-9017X issued by the Chandra X-ray Observatory Center, which is operated by the Smithsonian Astrophysical Observatory for and on behalf of the National Aeronautics Space Administration under contract NAS8-03060. S.J.W. is supported by NASA contract NAS8-03060 (Chandra). The NRAO VLA data are from programs  S7874 (NGC1333) and S9056 (IC 348).}
%VLA program IDs are S7874 (NGC1333) and S90564 (IC 348).  Need to include this in the paper for NRAO page charge support.

{\it Facilities:} \facility{CXO (ACIS), VLA}

\bibliographystyle{aa} % style aa.bst
\bibliography{bib_ic348} % your references Yourfile.bib

\begin{thebibliography}{47}
\expandafter\ifx\csname natexlab\endcsname\relax\def\natexlab#1{#1}\fi

\bibitem[{{Allen} {et~al.}(2004){Allen}, {Jerius}, \& {Gaetz}}]{all04}
{Allen}, C., {Jerius}, D.~H., \& {Gaetz}, T.~J. 2004, in Society of
  Photo-Optical Instrumentation Engineers (SPIE) Conference Series, Vol. 5165,
  Society of Photo-Optical Instrumentation Engineers (SPIE) Conference Series,
  ed. {K.~A.~Flanagan \& O.~H.~W.~Siegmund}, 423--432

\bibitem[{{Andr{\'e}}(1987)}]{and87}
{Andr{\'e}}, P. 1987, in Protostars and Molecular Clouds, ed. {T.~Montmerle \&
  C.~Bertout}, 143

\bibitem[{{Andr\'e}(1996)}]{and96}
{Andr\'e}, P. 1996, in Astronomical Society of the Pacific Conference Series,
  Vol.~93, Radio Emission from the Stars and the Sun, ed. {A.~R.~Taylor \&
  J.~M.~Paredes}, 273

\bibitem[{{Andr\'e} {et~al.}(1993){Andr\'e}, {Ward-Thompson}, \&
  {Barsony}}]{and93}
{Andr\'e}, P., {Ward-Thompson}, D., \& {Barsony}, M. 1993, \apj, 406, 122

\bibitem[{{Avila} {et~al.}(2001){Avila}, {Rodr{\'{\i}}guez}, \&
  {Curiel}}]{avi01}
{Avila}, R., {Rodr{\'{\i}}guez}, L.~F., \& {Curiel}, S. 2001, Revista Mexicana
  de Astronomia y Astrofisica, 37, 201

\bibitem[{{Bally} {et~al.}(2008){Bally}, {Walawender}, {Johnstone}, {Kirk}, \&
  {Goodman}}]{bal08}
{Bally}, J., {Walawender}, J., {Johnstone}, D., {Kirk}, H., \& {Goodman}, A.
  2008, {The Perseus Cloud}, ed. {Reipurth, B.}, 308

\bibitem[{{Benz} {et~al.}(2005){Benz}, {Grigis}, {Csillaghy}, \&
  {Saint-Hilaire}}]{ben05}
{Benz}, A.~O., {Grigis}, P.~C., {Csillaghy}, A., \& {Saint-Hilaire}, P. 2005,
  \solphys, 226, 121

\bibitem[{{Benz} \& {G\"udel}(1994)}]{ben94}
{Benz}, A.~O. \& {G\"udel}, M. 1994, \aap, 285, 621

\bibitem[{{Benz} {et~al.}(2006){Benz}, {Perret}, {Saint-Hilaire}, \&
  {Zlobec}}]{ben06}
{Benz}, A.~O., {Perret}, H., {Saint-Hilaire}, P., \& {Zlobec}, P. 2006,
  Advances in Space Research, 38, 951

\bibitem[{{Bower} {et~al.}(2003){Bower}, {Plambeck}, {Bolatto}, {McCrady},
  {Graham}, {de Pater}, {Liu}, \& {Baganoff}}]{bow03}
{Bower}, G.~C., {Plambeck}, R.~L., {Bolatto}, A., {et~al.} 2003, \apj, 598,
  1140

\bibitem[{{Chandler} \& {Richer}(2000)}]{cha00}
{Chandler}, C.~J. \& {Richer}, J.~S. 2000, \apj, 530, 851

\bibitem[{{Chen} {et~al.}(2009){Chen}, {Launhardt}, \& {Henning}}]{che09}
{Chen}, X., {Launhardt}, R., \& {Henning}, T. 2009, \apj, 691, 1729

\bibitem[{{Dulk}(1985)}]{dul85}
{Dulk}, G.~A. 1985, \araa, 23, 169

\bibitem[{{Enoch} {et~al.}(2009){Enoch}, {Evans}, {Sargent}, \&
  {Glenn}}]{eno09}
{Enoch}, M.~L., {Evans}, N.~J., {Sargent}, A.~I., \& {Glenn}, J. 2009, \apj,
  692, 973

\bibitem[{{Feigelson} {et~al.}(1998){Feigelson}, {Carkner}, \&
  {Wilking}}]{fei98}
{Feigelson}, E.~D., {Carkner}, L., \& {Wilking}, B.~A. 1998, \apjl, 494, L215

\bibitem[{{Feigelson} \& {Montmerle}(1999)}]{fei99}
{Feigelson}, E.~D. \& {Montmerle}, T. 1999, \araa, 37, 363

\bibitem[{{Feigelson} {et~al.}(1994){Feigelson}, {Welty}, {Imhoff}, {Hall},
  {Etzel}, {Phillips}, \& {Lonsdale}}]{fei94}
{Feigelson}, E.~D., {Welty}, A.~D., {Imhoff}, C., {et~al.} 1994, \apj, 432, 373

\bibitem[{{Forbrich} \& {Preibisch}(2007)}]{fop07}
{Forbrich}, J. \& {Preibisch}, T. 2007, \aap, 475, 959

\bibitem[{{Forbrich} {et~al.}(2006){Forbrich}, {Preibisch}, \&
  {Menten}}]{for06}
{Forbrich}, J., {Preibisch}, T., \& {Menten}, K.~M. 2006, \aap, 446, 155

\bibitem[{{Forbrich} {et~al.}(2007){Forbrich}, {Preibisch}, {Menten},
  {Neuh{\"a}user}, {Walter}, {Tamura}, {Matsunaga}, {Kusakabe}, {Nakajima},
  {Brandeker}, {Fornasier}, {Posselt}, {Tachihara}, \& {Broeg}}]{for07}
{Forbrich}, J., {Preibisch}, T., {Menten}, K.~M., {et~al.} 2007, \aap, 464,
  1003

\bibitem[{{Forbrich} {et~al.}(2011){Forbrich}, {Wolk}, {G\"udel}, {Benz},
  {Osten}, {Linsky}, {McLean}, {Loinard}, \& {Berger}}]{for11}
{Forbrich}, J., {Wolk}, S.~J., {G\"udel}, M., {et~al.} 2011, in Proceedings of
  Cool Stars 16, Vol. 000, Astronomical Society of the Pacific Conference
  Series, ed. {C. Johns-Krull, M. Browning, \& Andrew West}, tbd,
  arXiv:1012.1626

\bibitem[{{Froebrich} {et~al.}(2003){Froebrich}, {Smith}, {Hodapp}, \&
  {Eisl{\"o}ffel}}]{fro03}
{Froebrich}, D., {Smith}, M.~D., {Hodapp}, K., \& {Eisl{\"o}ffel}, J. 2003,
  \mnras, 346, 163

\bibitem[{{Gagn{\'e}} {et~al.}(2004){Gagn{\'e}}, {Skinner}, \&
  {Daniel}}]{gag04}
{Gagn{\'e}}, M., {Skinner}, S.~L., \& {Daniel}, K.~J. 2004, \apj, 613, 393

\bibitem[{{Getman} {et~al.}(2002){Getman}, {Feigelson}, {Townsley}, {Bally},
  {Lada}, \& {Reipurth}}]{get02}
{Getman}, K.~V., {Feigelson}, E.~D., {Townsley}, L., {et~al.} 2002, \apj, 575,
  354

\bibitem[{{G{\"u}del}(2002)}]{gue02}
{G{\"u}del}, M. 2002, \araa, 40, 217

\bibitem[{{G\"udel} \& {Benz}(1993)}]{gue93}
{G\"udel}, M. \& {Benz}, A.~O. 1993, \apjl, 405, L63

\bibitem[{{Guedel} {et~al.}(1997){Guedel}, {Guinan}, \& {Skinner}}]{gue97}
{Guedel}, M., {Guinan}, E.~F., \& {Skinner}, S.~L. 1997, \apj, 483, 947

\bibitem[{{Guenther} {et~al.}(2000){Guenther}, {Stelzer}, {Neuh{\"a}user},
  {Hillwig}, {Durisen}, {Menten}, {Greimel}, {Barwig}, {Englhauser}, \&
  {Robb}}]{gue00}
{Guenther}, E.~W., {Stelzer}, B., {Neuh{\"a}user}, R., {et~al.} 2000, \aap,
  357, 206

\bibitem[{{Gutermuth} {et~al.}(2008){Gutermuth}, {Myers}, {Megeath}, {Allen},
  {Pipher}, {Muzerolle}, {Porras}, {Winston}, \& {Fazio}}]{gut08}
{Gutermuth}, R.~A., {Myers}, P.~C., {Megeath}, S.~T., {et~al.} 2008, \apj, 674,
  336

\bibitem[{{Herbst}(2008)}]{her08}
{Herbst}, W. 2008, {Star Formation in IC 348}, ed. {Reipurth, B.}, 372

\bibitem[{{Hirota} {et~al.}(2008){Hirota}, {Bushimata}, {Choi}, {Honma},
  {Imai}, {Iwadate}, {Jike}, {Kameya}, {Kamohara}, {Kan-Ya}, {Kawaguchi},
  {Kijima}, {Kobayashi}, {Kuji}, {Kurayama}, {Manabe}, {Miyaji}, {Nagayama},
  {Nakagawa}, {Oh}, {Omodaka}, {Oyama}, {Sakai}, {Sasao}, {Sato}, {Shibata},
  {Tamura}, \& {Yamashita}}]{hir08}
{Hirota}, T., {Bushimata}, T., {Choi}, Y.~K., {et~al.} 2008, \pasj, 60, 37

\bibitem[{{Lada}(1987)}]{lad87}
{Lada}, C.~J. 1987, in IAU Symposium, Vol. 115, Star Forming Regions, ed.
  {M.~Peimbert \& J.~Jugaku}, 1--17

\bibitem[{{Lada} {et~al.}(2006){Lada}, {Muench}, {Luhman}, {Allen}, {Hartmann},
  {Megeath}, {Myers}, {Fazio}, {Wood}, {Muzerolle}, {Rieke}, {Siegler}, \&
  {Young}}]{lad06}
{Lada}, C.~J., {Muench}, A.~A., {Luhman}, K.~L., {et~al.} 2006, \aj, 131, 1574

\bibitem[{{Lombardi} {et~al.}(2010){Lombardi}, {Lada}, \& {Alves}}]{lom10}
{Lombardi}, M., {Lada}, C.~J., \& {Alves}, J. 2010, \aap, 512, A67

\bibitem[{{Muench} {et~al.}(2007){Muench}, {Lada}, {Luhman}, {Muzerolle}, \&
  {Young}}]{mue07}
{Muench}, A.~A., {Lada}, C.~J., {Luhman}, K.~L., {Muzerolle}, J., \& {Young},
  E. 2007, \aj, 134, 411

\bibitem[{{Mukai}(1993)}]{muk93}
{Mukai}, K. 1993, Legacy, 3, 21

\bibitem[{{Osten} \& {Wolk}(2009)}]{ost09}
{Osten}, R.~A. \& {Wolk}, S.~J. 2009, \apj, 691, 1128

\bibitem[{{Preibisch} \& {Feigelson}(2005)}]{pre05}
{Preibisch}, T. \& {Feigelson}, E.~D. 2005, \apjs, 160, 390

\bibitem[{{Rodr{\'{\i}}guez} {et~al.}(1997){Rodr{\'{\i}}guez}, {Anglada}, \&
  {Curiel}}]{rod97}
{Rodr{\'{\i}}guez}, L.~F., {Anglada}, G., \& {Curiel}, S. 1997, \apjl, 480,
  L125

\bibitem[{{Rodr{\'{\i}}guez} {et~al.}(1999){Rodr{\'{\i}}guez}, {Anglada}, \&
  {Curiel}}]{rod99}
{Rodr{\'{\i}}guez}, L.~F., {Anglada}, G., \& {Curiel}, S. 1999, \apjs, 125, 427

\bibitem[{{Shirley} {et~al.}(2007){Shirley}, {Claussen}, {Bourke}, {Young}, \&
  {Blake}}]{shi07}
{Shirley}, Y.~L., {Claussen}, M.~J., {Bourke}, T.~L., {Young}, C.~H., \&
  {Blake}, G.~A. 2007, \apj, 667, 329

\bibitem[{{Walawender} {et~al.}(2008){Walawender}, {Bally}, {Francesco},
  {J{\o}rgensen}, \& {Getman}}]{wal08}
{Walawender}, J., {Bally}, J., {Francesco}, J.~D., {J{\o}rgensen}, J., \&
  {Getman}, K. 2008, {NGC 1333: A Nearby Burst of Star Formation}, ed.
  {Reipurth, B.}, 346

\bibitem[{{White} {et~al.}(1992){White}, {Pallavicini}, \& {Kundu}}]{whi92}
{White}, S.~M., {Pallavicini}, R., \& {Kundu}, M.~R. 1992, \aap, 259, 149

\bibitem[{{Wilking} {et~al.}(2008){Wilking}, {Gagn{\'e}}, \& {Allen}}]{wil08}
{Wilking}, B.~A., {Gagn{\'e}}, M., \& {Allen}, L.~E. 2008, {Star Formation in
  the {$\rho$} Ophiuchi Molecular Cloud}, ed. {Reipurth, B.}, 351

\bibitem[{{Wilking} {et~al.}(2004){Wilking}, {Meyer}, {Greene}, {Mikhail}, \&
  {Carlson}}]{wil04}
{Wilking}, B.~A., {Meyer}, M.~R., {Greene}, T.~P., {Mikhail}, A., \& {Carlson},
  G. 2004, \aj, 127, 1131

\bibitem[{{Windhorst} {et~al.}(1993){Windhorst}, {Fomalont}, {Partridge}, \&
  {Lowenthal}}]{win93}
{Windhorst}, R.~A., {Fomalont}, E.~B., {Partridge}, R.~B., \& {Lowenthal},
  J.~D. 1993, \apj, 405, 498

\bibitem[{{Winston} {et~al.}(2010){Winston}, {Megeath}, {Wolk}, {Spitzbart},
  {Gutermuth}, {Allen}, {Hernandez}, {Covey}, {Muzerolle}, {Hora}, {Myers}, \&
  {Fazio}}]{win10}
{Winston}, E., {Megeath}, S.~T., {Wolk}, S.~J., {et~al.} 2010, \aj, 140, 266

\end{thebibliography}

\end{document}